\documentclass[aip,twocolumn,reprint,superscriptaddress]{revtex4-1}
\usepackage{amsmath}
\usepackage{amssymb}
\usepackage[latin9]{inputenc}
\usepackage{graphicx}
\usepackage{hyperref}
\usepackage{physics}
\usepackage{multibib}

\usepackage{graphicx}
\usepackage[color= blue!20]{todonotes}
\usepackage[normalem]{ulem}
\usepackage{color}
\usepackage{ulem}
\usepackage[symbol]{footmisc}

\usepackage{xcolor}
\hypersetup{
        colorlinks,
        linkcolor={blue!50!black},
        citecolor={blue!50!black}
}
\begin{document}
        
\addtocontents{toc}{\protect\setcounter{tocdepth}{0}}

\onecolumngrid

\noindent\textbf{\textsf{\Large Generation of Large Amplitude Phonon States in Quantum Acoustics}}

\normalsize
\vspace{.3cm}
\noindent\textsf{Clinton~A.~Potts$^{1\dagger}$, Wilfred~J.M.~Franse$^{1}$, Victor~Augusto~S.V.~Bittencourt$^{2}$, Anja~Metelmann$^{2,3,4}$ and Gary~A.~Steele$^{1*}$}

\vspace{.2cm}
\noindent\textit{$^1$Kavli Institute of Nanoscience, Delft University of Technology, PO Box 5046, 2600 GA Delft, The Netherlands\\$^2$Institut de Science et d'Ing\'{e}nierie Supramol\'{e}culaires (ISIS, UMR7006), Universit\'{e} de Strasbourg and CNRS, 67000 Strasbourg, France\\$^3$Institute for Theory of Condensed Matter, Karlsruhe Institute of Technology, 76131 Karlsruhe, Germany\\$^4$Institute for Quantum Materials and Technology, Karlsruhe Institute of Technology, 76344 Eggenstein-Leopoldshafen, Germany}

\vspace{.2cm}
\noindent$^{\dagger}$Email: clinton.potts@nbi.ku.dk \\
\noindent$^*$Email: g.a.steele@tudelft.nl

\vspace{.5cm}

\date{\today}

{\addtolength{\leftskip}{10 mm}
\addtolength{\rightskip}{10 mm}

The development of quantum acoustics has enabled the cooling of mechanical objects to their quantum ground state, generation of mechanical Fock-states, and Schr\"{o}dinger cat states. Such demonstrations have made mechanical resonators attractive candidates for quantum information processing, metrology, and macroscopic tests of quantum mechanics. However, generating large-amplitude phonon states in quantum acoustic systems has been elusive. In this work, a single superconducting qubit coupled to a high-overtone bulk acoustic resonator is used to generate a large phonon population in an acoustic mode of a high-overtone resonator. We observe extended ringdowns of the qubit, confirming the generation of a large amplitude phonon state, and also observe an upper threshold behavior, a consequence of phonon quenching predicted by our model. This work provides a key tool for generating arbitrary phonon states in circuit quantum acoustodynamics, which is important for fundamental and quantum information applications.
}
\vspace{.5cm}

\twocolumngrid

\section*{Introduction}

The prospect of controlling and manipulating phonons at the quantum level has sparked considerable interest, in particular for applications of non-classical phonon states for quantum sensing \cite{fadel2023probing,pino2018chip,zivari2022non}. The latter might enable tests of gravity effects in quantum mechanics, a long-pursued milestone in physics \cite{penrose1996gravity,diosi1987universal, Gely_Superconducting_2021}. In this context, circuit quantum acoustodynamics (cQAD)\cite{chu2017quantum} has emerged as a promising platform for controlling phonons at the quantum regime. In such systems, a single superconducting qubit is coupled to an acoustic phonon mode, enabling the use of the well-developed circuit quantum electrodynamics toolkit \cite{blais2021circuit,krantz2019quantum,hofheinz2009synthesizing}  for manipulating qubits as a way to control phonons \cite{undershute2025decoherence,franse2024high,manenti2017circuit,jiang2023thin}. As a consequence, cQAD has seen rapid success in the generation of mechanical states including the generation of Fock states \cite{chu2018creation} and Schr\"{o}dinger cat states \cite{bild2023schrodinger}. Moreover, cQAD has prospects for developing novel hybrid quantum systems \cite{chu2020perspective, valimaa2022multiphonon, hann2019hardware}, allowing further integration of the system with other hybrid architectures \cite{clerk2020hybrid}.

In this article, we demonstrate the generation of a large phonon population in cQAD using a scheme similar to single-atom lasers \cite{mckeever2003experimental, dubin2010quantum,an1994microlaser}. A single superconducting qubit couples to a bulk on-chip phonon mode, generating a large amplitude phonon state through the process of stimulated emission. The phonon state amplitude is confirmed through a dramatically extended ringdown of the superconducting qubit excited by the phonon state. We also observe a unique predicted feature of our model, an upper threshold, which is a consequence of population quenching, a phenomenon also present in single-atom lasers \cite{dubin2010quantum, mu1992one, ashhab2009single}. Close to such an upper threshold, the phonon mode experiences a reduced linewidth, indicating a qubit-induced amplification. Our experimental observations agree with the theoretical and numerical models we use to characterize the system. 

\begin{figure}
    \centering
    \includegraphics[width = 0.45\textwidth]{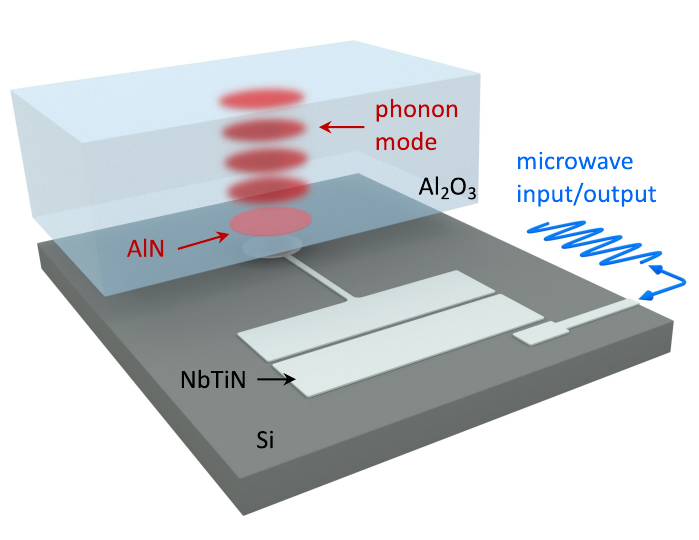}
    \caption{\textbf{Schematic of the on-chip $\hbar$BAR device.} Rendering of the $\hbar$BAR device. The $\hbar$BAR device comprises two chips bonded in a flip-chip orientation. The top chip is 650 $\mu$m of sapphire, hosts the high-overtone bulk acoustic wave resonances (HBAR) modes (red), and is coupled to the superconducting antenna using an aluminum nitride pad (red). The pocket-style transmon qubit (silver) is fabricated from niobium titanium nitride on the bottom silicon chip and coupled to the feedline via an on-chip readout resonator (blue). Art produced by Enrique Sahagun \cite{Scixel}.}
    \label{Figure1}
\end{figure}

Finally, the phonon mode's bulk nature may allow the integration of color centers or quantum dots, enabling strain engineering or coherent mechanical driving for future hybrid quantum systems. Moreover, the coherent state generated here may be used as an efficient displacement pulse for generating large Schr\"{o}dinger cat states \cite{bild2023schrodinger} or squeezed states \cite{marti2023quantum}. Generating large mechanical cat states has exciting potential applications for future macroscopic tests of quantum mechanics \cite{Schrinski_Macroscopic_2023}.

\section*{Results}

\subsection*{Experimental Setup}

Our device comprises a flip-chip $\hbar$BAR architecture with two bonded device chips \cite{franse2024high,bild2023schrodinger,von2023engineering,von2022parity,chu2018creation,chu2017quantum}. We have implemented a fully on-chip integration, which can multiplex different devices on a single silicon chip \cite{Alpo_2021_2Dhbar,crump2023coupling}. A single feedline is coupled to the individual superconducting fixed-frequency transmon qubits. The transmon qubits were fabricated from niobium-titanium nitride for the bulk structures with aluminum Josephson junctions. A sapphire substrate, $650$ $\mu$m thick, was positioned above each transmon qubit and bonded to the silicon substrate, see Fig.~\ref{Figure1}. The sapphire chip supports a set of longitudinal high-overtone bulk acoustic wave resonances (HBARs) separated by a free spectral range of $8.54$ MHz. The electric field of a transmon qubit coherently couples to the strain of an HBAR acoustic mode via a disk of piezoelectric aluminum nitride patterned on the sapphire. Each qubit on the chip is nearly resonant with an HBAR mode of interest; in such a way, a pair of qubit-HBAR modes behaves like a single atom coupled to a phononic mode. The qubits were read out via on-chip microwave resonators using standard circuit quantum electrodynamics techniques \cite{krantz2019quantum,reed2010high}. Our device is similar to those used in previous work generating mechanical Schr\"{o}dinger cat states \cite{bild2023schrodinger} and for circuit quantum acoustodynamics \cite{chu2017quantum}.

\begin{figure*}
    \centering
    \includegraphics[width = 0.95\textwidth]{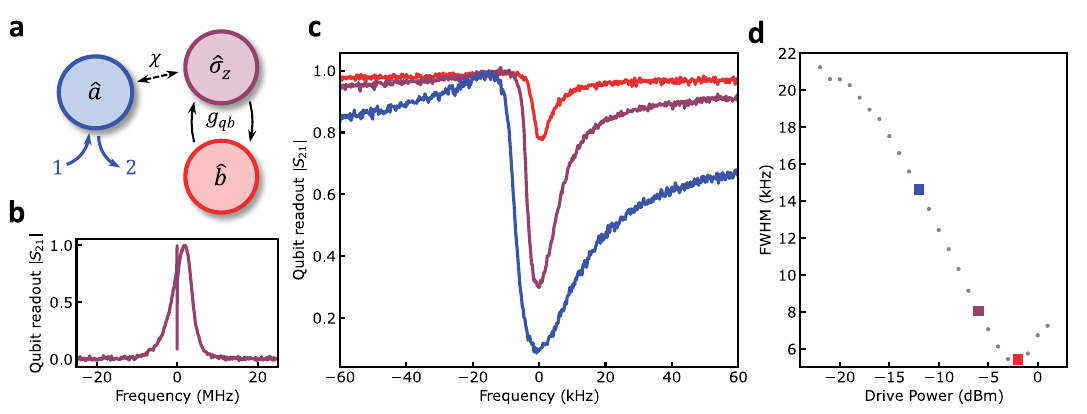}
    \caption{\textbf{Qubit-induced acoustic linewidth narrowing.} \textbf{a} Definition of modes and the coupling rates between the modes. The readout resonator $\hat{a}$ has input-output modes labelled $1$ and $2$, and is dispersively coupled to the qubit $\hat{\sigma}_{\rm z}$ with a rate $\chi$. The mechanical mode is labeled $\hat{b}$ and has a qubit-phonon coupling rate $g_{\rm qb}$. \textbf{b} Measured two-tone spectroscopy for drive power of -12.0 dBm set at room temperature. \textbf{c} Measured phonon-induced transparency window as a function of qubit drive power. Starting from the lowest curve, drive powers are -12.0, -6.0, and -2.0 dBm set at room temperature, respectively. With increasing power, two features can be noticed. The transparency window reduces in depth, and the full-width half-maximum linewidth narrows. \textbf{d} Extracted experimental full-width half-maximum of the transparency window as a function of qubit drive power. Colored points match the corresponding curves in \textbf{c}.}
    \label{Figure2}
\end{figure*}

Using the dispersive shift of a coupled linear readout resonator, we can measure the steady-state qubit population \cite{krantz2019quantum}. In the limit where the qubit and the readout resonator are far detuned in frequency, the qubit-cavity Hamiltonian can be written as:
\begin{equation}
    \hat{\mathcal{H}}/\hbar = \omega_{\rm r}\hat{a}^{\dagger}\hat{a} + \frac{1}{2}\omega_{\rm q}\hat{\sigma}_{\rm z} + \chi \hat{\sigma}_{\rm z} \hat{a}^{\dagger}\hat{a},
\end{equation}
where $\omega_{\rm r,q}$ are the readout and qubit frequency, $\hat{a}^{(\dagger)}$ is the photon annihilation (creation) operator, $\hat{\sigma}_{\rm z}$ is the qubit Pauli-z operator, and $\chi$ is the qubit-state dependent frequency shift of the readout resonator. The coupling between the phonon and the qubit is described by a resonant Jaynes-Cummings interaction \cite{chu2017quantum}, given by the Hamiltonian:
\begin{equation}
    \hat{\mathcal{H}}_{\rm int}/\hbar = g_{\rm qb}( \hat{\sigma}_+ \hat{b} + \hat{\sigma}_- \hat{b}^{\dagger}),
\end{equation}
where $g_{\rm qb}$ is the coupling rate between the qubit and the phonon mode, $\hat{\sigma}_{\pm}$ are the qubit raising and lowering operators, and $\hat{b}^{(\dagger)}$ is the phonon annihilation (creation) operator; see Fig.~\ref{Figure2}(a). 

The qubit's state was measured by applying a weak probe tone on resonance with the readout resonator and monitoring the transmitted signal as a second tone was swept near the qubit frequency. The transmission spectrum at the readout frequency directly maps to the qubit occupation $\langle \hat{\sigma}_{+}\hat{\sigma}_- \rangle$; see Fig.~\ref{Figure2}(b). The asymmetry of the qubit spectrum is due to the finite photon population within the readout resonator \cite{gambetta2006qubit,lachance2017resolving} and is well described by our theoretical model. Moreover, the narrow transparency window within the qubit spectrum results from the weak hybridization between the qubit and the phonon mode. In this work, the qubit and the phonon frequency were detuned by approximately $3.3$ MHz. See the discussion in the supplementary text for full details. However, as will be discussed below, coherent energy exchange between the qubit and phonon modes does not provide a clear picture of the dynamics at high drive powers. 

\begin{figure}
    \centering
    \includegraphics[width = 0.45\textwidth]{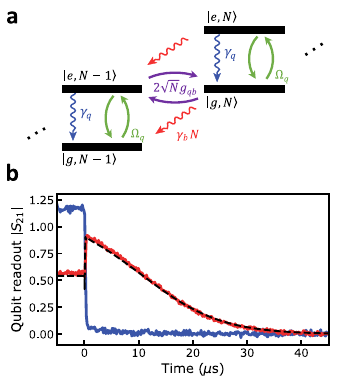}
    \caption{\textbf{Large amplitude phonon states observed through qubit ringdown dynamics.} \textbf{a} Schematic of the energy levels for the weakly hybridized phonon-qubit system. The qubit acts as an artificial two-level atom, with ground state $\vert g \rangle$ and excited state $\vert e \rangle$. A coherent drive of strength $\Omega_{\rm q}$ drives the qubit between its ground and excited state. The qubit-mechanical couple at a rate $g_{\rm qb}$ and a qubit and mechanics decay at a rate $\gamma_{\rm q}$ and $\gamma_{\rm b}$, respectively. The rate at which excitation transfer processes $\vert e, N-1 \rangle \rightarrow \vert g, N \rangle$ occur scales as $\sqrt{N}$, where $N$ is the total number of excitations, while the phonon relaxation scales linearly with $N$. The qubit is rapidly re-excited for strong pump powers, resulting in a build-up of phonon excitations. \textbf{b} Measured qubit ringdown for a qubit drive of -3.0 dBm.  Blue line: the qubit drive is detuned from the HBAR by $250$ kHz with the phonon mode in the non-lasing state, decaying on a time scale of the $\sim 200$ ns decay of the qubit. Red line: The qubit drive is tuned directly on resonance with the HBAR mode, exciting it into the lasing state, exhibiting a dramatically longer, non-exponential decay due to re-excitation from the coherently excited phonon mode. The master equation simulation (dashed black line) is plotted over the data.}
    \label{Figure3}
\end{figure}

\subsection*{Phonon Generation}

Three distinct features can be observed within the two-tone spectrum when increasing the qubit drive power. First, the qubit linewidth is power-broadened \cite{schuster2005ac,gambetta2006qubit}. Large qubit drive powers increase the qubit decay rate due to increased stimulated emission. Less intuitive is the gradual disappearance and narrowing of the phonon-induced transparency window; see Fig.~\ref{Figure2}(c). The total linewidth of the transparency window is proportional to the phonon-qubit cooperativity and demonstrates an inverse dependence on the drive power; see Fig.~\ref{Figure2}(d). Such behavior can not be entirely explained by the power broadening of the qubit linewidth. At powers closer to the minimum of the curve Fig.~\ref{Figure2}(d), the narrowing of the transparency window linewidth can be associated with an amplification process triggered by the coupling to the driven qubit. Above such an upper threshold, the linewidth increases, a feature that can not be attributed to a suppression of the inverse Purcell effect due to the qubit broadening. These features can be understood by considering the schematic shown in Fig.~\ref{Figure3}(a).

The build-up of the phonon population can be understood by considering the different processes through which excitations can be transferred between the phonon and the qubit. At low driving powers, the linewidth broadening of the qubit due to the microwave drive is small, and the phonon mode and qubit are weakly interacting. The weak interaction combined with the microwave drive initiates the buildup of the phonon population. A balance between the rate at which the qubit is excited and de-excited, the rate at which excitations are transferred between qubit and phonon mode and the stochastic phonon losses yields a steady-state occupation of the phonon mode. For low drive powers and thus slow qubit excitation rates, phonon loss dominates, preventing the build-up of the phonon population. As the qubit excitation rate increases, transitions between states of the Jaynes-Cummings ladder with higher phonon numbers can be efficiently driven. Since the rate at which excitations are transferred between states of the Jaynes-Cummings ladder depends on the phonon number of the involved states, as the average population of the phonon mode increases, the rate at which excitations are transferred correspondingly increases; this is the origin of acoustic stimulated emission \cite{hofheinz2008generation}.

In the steady state, this process, balanced by stochastic phonon losses, results in a large phonon population. Such processes also provide an intuitive understanding of the reduced visibility of the transparency window. As the drive power increases, the linewidth of the qubit becomes increasingly broad, and the coupling between the qubit and phonon is not strong enough to overcome the qubit loss; therefore, the qubit spectroscopy provides less information about the phonon mode.

At the highest qubit drive powers, the qubit linewidth has been increased such that the rapid decay of the qubit results in a quenching of phonon generation \cite{mu1992one}. This results in an upper threshold above which the phonon mode is no longer effectively excited. Above this threshold, the qubit undergoes rapid Rabi oscillations due to the strong microwave drive and, therefore, cannot exchange excitations with the phonon mode, reducing the phonon amplitude. The buildup of the phonon population, primarily due to stimulated emission from a single quantum emitter, including a reduced phonon mode linewidth and the upper threshold behavior, is similar to a single-atom laser. It should also be noted that the phonon mode statistics are no longer predicted to be that of a coherent state above the upper threshold. This process has been described previously in the context of single-atom photon lasers \cite{mckeever2003experimental} and is captured by our theoretical description; see the supplementary text.

\begin{figure*}
    \centering
    \includegraphics[width = 0.95\textwidth]{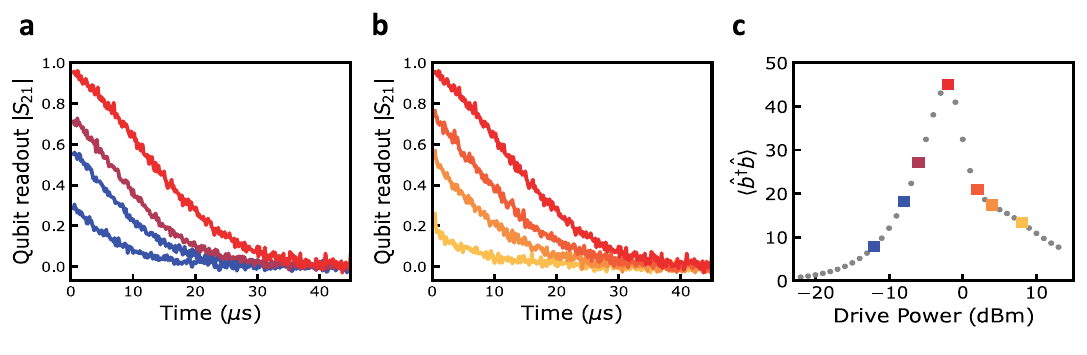}
    \caption{\textbf{Unique signatures of predicted upper threshold.} \textbf{a} Measured gated qubit ringdown for a qubit drive of -12.0, -8.0, -6.0, and -2.0  dBm. With increasing drive power, the gated qubit ringdowns increase in amplitude and duration as the phonon mode population increases. \textbf{b} Measured gated qubit ringdown for a qubit drive of -2.0, 2.0, 4.0, and 8.0  dBm. The qubit drive is tuned directly on resonance with the HBAR mode for all measurements. With increasing drive power, the gated qubit ringdowns decrease in amplitude and duration as the phonon mode population decreases above the self-quenching threshold. For all measurements in \textbf{a} and \textbf{b}, the qubit drive is tuned directly on resonance with the HBAR mode. \textbf{c} Simulated phonon state population $\langle \hat{b}^{\dagger}\hat{b} \rangle$ as a function of qubit drive power. The phonon population is at its maximum at a drive power of approximately -2.0 dBm and decreases for drive powers above the upper threshold. The colored data points indicate the corresponding trace colors in \textbf{a} and \textbf{b}. }
    \label{Figure4}
\end{figure*}

\subsection*{Gated Two-Tone Spectroscopy}

Directly probing the phonon mode is not possible in the current experimental configuration since the readout is performed via the two-level system and not through a propagating photon mode \cite{kuang2023nonlinear,behrl2023phonon}. Direct measurements of the Rabi oscillations between the qubit and phonon state have been previously used to measure Fock-states in $\hbar$BAR devices \cite{chu2018creation}. However, this data would not be possible due to the short lifetime of our qubit and the multiplicity of simultaneous Rabi oscillation frequencies between two Fock states is given by $2g_{\rm N} = 2 \sqrt{N}g_{\rm qb}$, which scale with the phonon Fock number $N$.\cite{chu2018creation} Instead, we rely on the mismatch between the decay rate of the phonon mode $\tau \sim 25$ $\mu$s and that of the qubit. Using gated two-tone ringdown measurements, we can distinguish pure qubit decay from qubit decay driven by a highly excited phonon state. If the phonon mode is highly excited---in the absence of an external drive---the coherent Jaynes-Cummings interaction will continually drive the qubit, resulting in an extended relaxation of the qubit population compared to its intrinsic relaxation rate. 

Gated two-tone measurements were performed, driving the qubit until the system reached its steady state; at this point, the drive was switched off using an RF switch. During the entire sequence, the frequency of the readout resonator is monitored using a vector network analyzer, averaging multiple traces triggered synchronously with the RF switch; see the supplementary text for more information. This measures the expectation value of the qubit population $\langle \hat{\sigma}_{+}\hat{\sigma}_- \rangle$ as a function of time, with a temporal resolution of $50$ ns.

First, the drive power was set near the peak of the lasing amplitude and was detuned $250$ kHz above the HBAR resonance. The gated two-tone measurement was performed, and the blue data points in Fig.~\ref{Figure3}(b) show the resulting time domain measurement and a ringdown on the order of $\sim 200$ ns. This corresponds to the intrinsic $T_1$ decay of the transmon qubit. A second measurement was performed at the same drive power, but the drive was tuned resonant with the HBAR. A ringdown on the order of $25$ $\mu$s is observed for this configuration, represented by the red data points in Fig.~\ref{Figure3}(b). The extended ringdown confirms the highly excited phonon amplitude of the mechanical state. When the qubit drive is switched off, the phonon mode can exchange excitations with the qubit, continually re-exciting the qubit until the phonon mode has decayed back to its ground state. Moreover, near the peak phonon amplitude, in contrast to a thermal state, the coherent state generated by the lasing results in a qubit ringdown that is not exponential; instead, the qubit ringdown is approximately linear. This feature is captured by our numerical model, the dashed curve in Fig.~\ref{Figure3}(b). Moreover, our model also captures the ring-up of the qubit, which is described in the supplementary information.

We can estimate the phonon population from our numerical model by fitting the spectroscopic and ringdown data. The estimated phonon population is shown in Fig.~\ref{Figure4}(c) as a function of qubit drive power. At a power of -2.0 dBm, the phonon population is nearly maximized, corresponding to the ringdown in Fig.~\ref{Figure4}(a,b), and the upper threshold is clearly visible as the phonon population rapidly reduces with increasing drive power. The upper threshold is experimentally confirmed by performing a set of gated ringdown measurements at a series of qubit drive powers. With increasing qubit drive power, the individual ringdown traces grow in amplitude and increase in duration, corresponding to the increasing phonon population, as shown in Fig.~\ref{Figure4}(a). At a drive power of -2.0 dBm, the phonon ringdown obtains its peak amplitude and duration, indicating a peak in the phonon population, shown in Fig.~\ref{Figure4}, which agrees with the minimum in the transparency window linewidth, shown in Fig.~\ref{Figure2}(b). Further increasing the qubit drive power beyond -2.0 dBm, the qubit ringdown decreases in amplitude and duration, a direct indication of the self-quenching, well described by our theoretical model and similar to what has been observed in single-atom lasers, see Fig.~\ref{Figure4}(b). The gated ringdown measurement demonstrates a clear upper threshold behavior and agrees with our numerical simulations and semi-classical analysis. We also notice that the qubit decay profile differs below and above the threshold for a given phonon population. Specifically, the decay is not exponential below the upper threshold, while well above the threshold, the decay is exponential. We associate such behavior with a change in the phonon state, which, according to our numerical simulations, is coherent below the upper threshold.

\section*{Discussion}

This article demonstrates the experimental realization and generation of large-amplitude phonon states in cQAD. Our experiment consists of a superconducting single-atom, realized using a transmon-style qubit coupled resonantly to an HBAR mode. When driving on resonance with the HBAR mode, the intrinsic non-linearity of the qubit-phonon coupling generates a highly excited phonon state. A key feature of this experiment is the driven two-level atom rather than the parametric instability driving phonon excitations. Moreover, the phonon mode is confined in a bulk longitudinal mode within a sapphire substrate. The bulk nature may allow the phonon mode's integration with additional on-chip architectures, such as color centers or quantum dots. 

Our results have demonstrated the successful generation of a large amplitude phonon state, and we have further demonstrated a counterintuitive feature predicted by our model, an upper threshold \cite{ashhab2009single}. The size of the coherent state achieved in this work was limited by both the qubit and phonon linewidth. However, the primary limiting factor was the phonon linewidth. Decreasing the decay rates will reduce the upper threshold power and increase the peak phonon amplitude. Future studies could include a linear probe or use a higher-order transition of the transmon to drive and measure the phonon statistics. Such a cQAD-compatible system promises to provide a highly coherent source of phonons, which have applications including sensing to quantum information processing and the generation of high-displacement, Schr\"{o}dinger cat states.

\section*{Methods}

\subsection*{Device Fabrication}

\subsection*{Qubit Chip}

The device fabrication starts with a 10x10mm chip 525 $\mu$m thick high resistivity $\langle100\rangle$ silicon deposited with $100$ nm of niobium-titanium nitride (NbTiN). The NbTiN film was deposited by the Dutch Institute for Space Research (SRON) following the process described in \cite{thoen2016superconducting}. A layer of photoresist (AR-P 6200.18, 4000 rpm) was patterned, exposed (EBPG 5200, 315 $\mu$m/cm$^2$) and developed (Pentylacetate, O-xylene, IPA) to form the bulk circuitry (transmon islands and coplanar waveguides). The exposed NbTiN was removed using a reactive ion etch (Sentech Etchlab 200, 13.5 sccm $\text{SF}_{\text{6}}$ + 5 sccm $\text{O}_{\text{2}}$, 55 W, 10 $\mu$bar) followed by an in-situ oxygen descum (50 sccm $\text{O}_{\text{2}}$,100 W, 10 $\mu$bar). After stripping the photoresist, a bilayer resist stack (MAA 8.5\% EL6, 2000 rpm and PMMA A6 950k, 1500 rpm; baked for three and five minutes at 180 $^{\circ}$C, respectively) was used for patterning the Josephson junctions (190 nm width). The junctions were patterned using e-beam lithography. The bilayer was developed using cold $\text{H}_{2}\text{O}$ : IPA (1:3) and cleaned afterwards with IPA. After cleaning the exposed silicon surface with an oxygen descum (200 sccm, 100 W) and acid clean (BoE(7:1):$\text{H}_{2}\text{O}$, 1:1), the chip was placed in an aluminum evaporator (Plassys MEB550). Double-angle shadow evaporation with intermediate in-situ oxidation was used to create Manhattan-style junctions. The aluminum was evaporated at a $35^{\circ}$ angle relative to the substrate at a rotational angle of $0^{\circ}$ and $90^{\circ}$. The top and bottom electrodes were  35 and 75 nm thick, respectively. After the first evaporation step, the aluminum was oxidized to create the Al$\text{O}_{\text{x}}$ tunnel barriers. Following the second evaporation step, a second oxidation step was performed to cap the junctions with a passivation layer. After performing liftoff in NMP, the qubit chip was finished.

\subsection*{HBAR Chip}

The HBAR chip started with double-side polished four-inch sapphire wafers with a 1 $\mu$m thick film of c-axis oriented AlN (Kyma technologies, AT.U.100.1000.B). The wafer was diced into 10x10mm chips for easier processing. A photoresist layer (AR-N 4450.10, 6000 rpm) was used to pattern circular regions, $250 \mu$m in diameter, to mask the AlN. A reactive ion etch in an Oxford 100 was performed to create AlN disks ($\text{Cl}_{2}\text{/BCl}_{3}\text{/Ar}$ at 4.0/26.0/10.0 sccm, 350 W ICP power, 70 W RF power). Following the reactive ion etch, the AlN layer has the proper shape but not the correct thickness. After stripping the photoresist, the chip was placed again inside the etcher to etch the AlN to $\sim900$ nm thickness.

\subsection*{Flip Chip}

Once fabrication on both chips was done, the HBAR chip was diced into 8x2 mm chips. The HBAR chip was then flipped on top of the qubit chip with the AlN layer facing down. Using probe needles, the AlN disks were aligned with the transmon antennas. Once aligned, the probe needles held down the chips in position while a tapered fiber was used to apply two-component epoxy (Loctite EA 3430) on the sides of the top chip; see the supplementary information. After the epoxy was cured, the chip was wire-bonded and installed onto the baseplate of the dilution refrigerator.

\subsection*{Measurement Setup}

\subsection*{Two-Tone Spectroscopy}

All measurements were performed within a dilution refrigerator operating at a base temperature $T \sim 20$ mK. A schematic of the dilution refrigerator setup and the room-temperature electronics are shown in the supplementary information. The device was mounted on the mixing chamber plate of the dilution refrigerator and connected to a set of coaxial cables. The device was measured in transmission, with the resonators coupled in a 'notch'-style geometry. The output signals went into a cryogenic HEMT (High Electron Mobility Transistor) amplifier (LNF-LNC4-8A), followed by additional room-temperature amplification (Miteq AFS3-04000800-07-10P-4). The input line was attenuated at each stage to reduce the electron temperature and the thermal radiation at the input port of our device. A total of 48 dB of attenuation was used, plus any additional attenuation from the coaxial cables. 

The two-tone spectroscopy was measured using a vector network analyzer (VNA). Port one and port three were combined using a directional coupler, with port three attached to the -20 dB coupling port. Port one was set into zero span mode and output a constant signal tuned on resonance with the Stark shifted readout resonator, $\omega_{\rm r}$, with an output power of -25 dBm. An additional 60 dB of attenuation was added to this signal before the directional coupler. Port 3 was used as a spectroscopic tone and was swept near the qubit frequency, and its power was varied throughout the experiment and had an additional 40 dB of attenuation. The combined signals from port one and there were then connected to the input line of the dilution refrigerator. 

The output from the dilution refrigerator was directly connected to port 4 of the VNA set in zero span mode at the readout resonator frequency $\omega_{\rm r}$. Two-tone spectroscopy was performed by slowly sweeping the qubit drive tone, ensuring the system has reached its steady state and monitoring the readout resonators transmission spectrum $S_{21}$.

\subsection*{Gated Two-Tone Spectroscopy}

For the gated two-tone measurement, port one of the VNA was set up just as in the two-tone measurement. An external signal generator generated the qubit drive tone. The qubit drive tone was passed through an RF switch before being combined with the readout tone using a directional coupler. The RF switch was triggered using a pulse generator at a 3 kHz repetition rate. The signal generator was set to a 22.5\% duty cycle, so the qubit drive was off for 75 $\mu$s per trace. The VNA was synchronously triggered by the signal generator, allowing for 65536 trace averages to be performed (maximum setting). Each trace-averaged measurement was repeated 75 additional times to improve the signal-to-noise ratio further.

\subsection*{Data Availability}

All data, analysis code, and measurement software are available in the manuscript or the supplementary material or are available at Zenodo https://doi.org/10.5281/zenodo.14810526.

\section*{Acknowledgments}

The authors thank Enrique Sahagun for the device rendering \cite{Scixel}. W.J.M.F and G.A.S. acknowledge support through the QUAKE project,  project number 680.92.18.04, of the research programme Natuurkunde Vrije Programma's of the Dutch Research Council (NWO). C.A.P. acknowledges the support of the Natural Sciences and Engineering Research Council of Canada (NSERC) (PDF-567689-2022). A.M and V.A.S.V.B. acknowledge financial support from the Contrat Triennal 2021-2023 Strasbourg Capitale Europeenne. 

\textbf{Author contributions} C.A.P. performed experiments, theoretical modelling, conceptualization, and wrote the manuscript with input from all authors. W.J.M.F. fabricated the device and performed experiments. V.A.S.V.B. performed theoretical modelling. A.M. provided supervision and funding acquisition. G.A.S. provided supervision, conceptualization and funding acquisition. 

\textbf{Ethics declarations}:

\textbf{Competing interests}: The authors declare no competing interests. 

\clearpage

\bibliography{apssamp}

\clearpage

\addtocontents{toc}{\protect\setcounter{tocdepth}{0}}

\onecolumngrid

\noindent\textbf{\textsf{\Large Supplementary Information: Large Amplitude Phonon States in Quantum Acoustics}}

\normalsize
\vspace{.3cm}
\noindent\textsf{C.A.~Potts$^{1,\dagger}$, W.J.M.~Franse$^{1}$, V.A.S.V.~Bittencourt$^{2}$, A.~Metelmann$^{2,3,4}$ and G.A.~Steele$^{1},^*$}

\vspace{.2cm}
\noindent\textit{$^1$Kavli Institute of Nanoscience, Delft University of Technology, PO Box 5046, 2600 GA Delft, The Netherlands\\$^2$ISIS (UMR 7006), Universit\'{e} de Strasbourg, 67000 Strasbourg, France\\$^3$Institute for Theory of Condensed Matter, Karlsruhe Institute of Technology, 76131 Karlsruhe, Germany\\$^4$Institute for Quantum Materials and Technology, Karlsruhe Institute of Technology, 76344 Eggenstein-Leopoldshafen, Germany}

\vspace{.2cm}
\noindent$^{\dagger}$Email: clinton.potts@nbi.ku.dk \\
\noindent$^*$Email: g.a.steele@tudelft.nl

\onecolumngrid

\setcounter{equation}{0}
\setcounter{figure}{0}
\setcounter{table}{0}
\renewcommand{\theequation}{S\arabic{equation}}
\renewcommand{\thefigure}{\arabic{figure}}
\renewcommand{\figurename}{SUPPLEMENTARY FIGURE}
\renewcommand{\tablename}{SUPPLEMENTARY TABLE}

\section{Supplementary Note 1: The Device}

\begin{figure*}[h!]
    \centering
    \includegraphics[width = 0.9\textwidth]{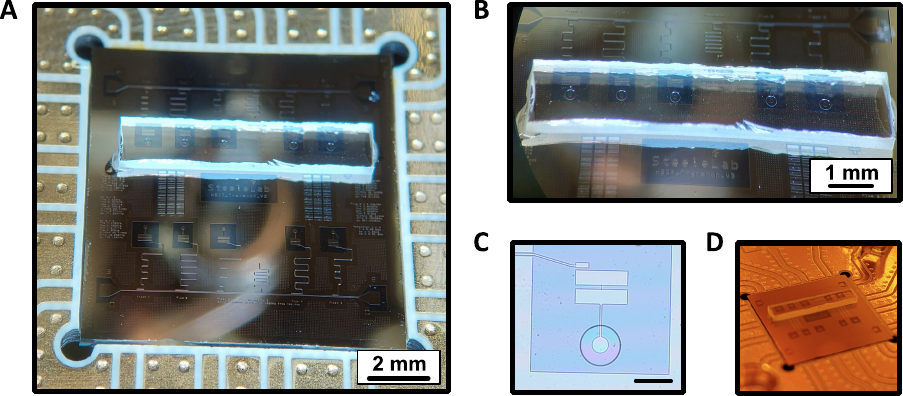}
    \caption{\textbf{Fabricated flip-chip HBAR Device.} (A) Optical micrograph of the assembled flip-chip device. The top feedline of qubits is fabricated with a flip-chip of sapphire, and the bottom feedline has no sapphire flip-chip as references. (B) Zoomed optical micrograph of the flip-chip assembly. (C) Optical micrograph of the qubit with the sapphire chip assembled on top. The overlap between the antenna and the $250 \mu$m aluminum nitride piezoelectric transducer is visible. The scale bar is $250 \mu$m. (D) Optical micrograph of the entire chip loaded in the printed circuit board.}
    \label{SIFig_Device}
\end{figure*}

\begin{figure}[h!]
    \includegraphics[width=0.8\textwidth]{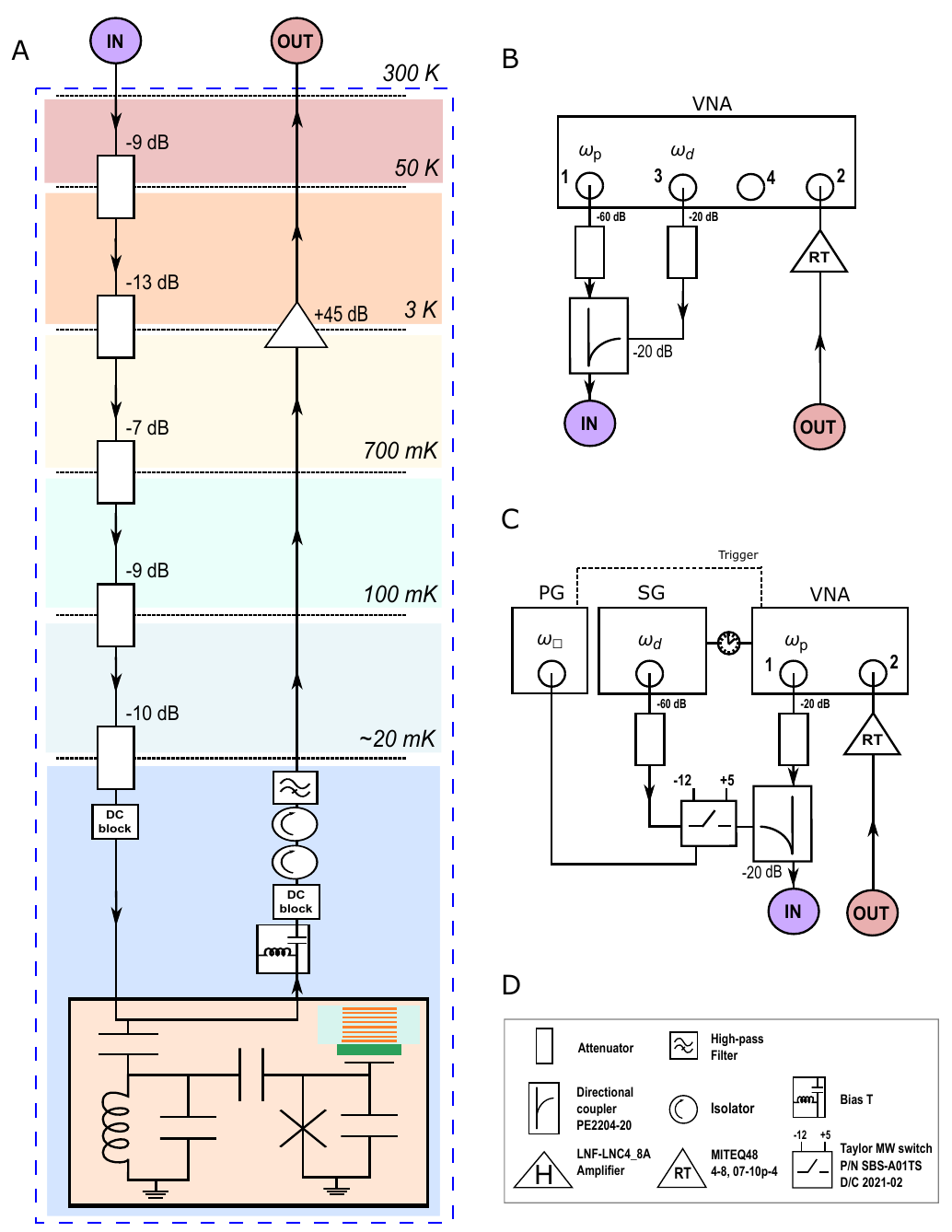}
    \caption{\textbf{Schematic of the measurement setup.} (A)  Dilution refrigerator wiring setup. Outside the refrigerator, we used two different setups. In (B), we show our Two-tone spectroscopy configuration used for qubit spectroscopy. Here, we sent a weak continuous wave tone (readout resonator probe) from the vector network analyzer (VNA) port one and a second continuous wave tone (qubit drive) from the VNA port 3. These two signals are combined using a directional coupler before entering the dilution refrigerator. The signal from the dilution refrigerator goes through a room-temperature amplifier before it goes to port 2 of the VNA. In (C), we show our "time domain" setup. Here, we replaced output port 3 of the VNA with a signal generator (SG) to provide the drive tone. A switch (\textit{Taylor MW switch}) together with a pulse generator (PG, \textit{Rigol DG1022}) is placed between the directional coupler and the signal generator. Adapted from \cite{franse2024high}}
    \label{Fig_Wiring_setup}
\end{figure}

\section{Supplementary Note 2: Theory}

\subsection{System Hamiltonian and Master Equation}

The system consists of a microwave cavity coupled to a transmon qubit which in turn is coupled to an HBAR phonon mode.  The transmon qubit can be considered, up to a good approximation, to be a two-level system. The readout cavity is driven two with coherent tones: one at a frequency $\omega_p$, which we call the probe tone, and one at a frequency $\omega_d$, which we call the drive tone. The experiment measures the coupling between the qubit and the phonon mode via two-tone spectroscopy performed via the cavity. The procedure consists of considering the cavity detuned from the qubit, setting the probe tone at the (shifted) cavity frequency and varying the drive tone close to the qubit Lamb-shifted frequency. The transmission of the cavity carries information about the qubit correlations $\langle \hat{\sigma}_z \rangle$. The phonon mode has a frequency close to the qubit frequency.

We model the system with the Hamiltonian
\begin{equation}
\begin{aligned}
\frac{\hat{\mathcal{H}}}{\hbar} &= \omega_{\rm r} \hat{a}^\dagger \hat{a} + \omega_{\rm b} \hat{b}^\dagger \hat{b} + \frac{\omega_{\rm q}}{2} \hat{\sigma}_{\rm z} + g_{\rm qc} (\hat{a} \hat{\sigma}_+ + \hat{a}^\dagger \hat{\sigma}_-) + g_{\rm qb} (\hat{b}^\dagger \hat{\sigma}_- + \hat{b} \hat{\sigma}_+)  \\
&+ \epsilon_{\rm d} (\hat{a} e^{i \omega_{\rm d} t} + \hat{a}^\dagger e^{-i \omega_{\rm d} t})+ \epsilon_{\rm p} (\hat{a} e^{i \omega_{\rm p} t} + \hat{a}^\dagger e^{-i \omega_{\rm p} t}),
\end{aligned}
\end{equation}
where $\hat{a}^{(\dagger)}$ are the annihilation (creation) operators for the readout cavity with frequency $\omega_{\rm r}$, $\hat{b}^{(\dagger)}$ are the annihilation (creation) operators for the HBAR mode with frequency $\omega_{\rm b}$, $\hat{\sigma}_{\rm z}$ is the qubit population operator with qubit frequency $\omega_{\rm q}$. The couplings are defined by the rates $g_{\rm qc}$ between the qubit and the readout cavity and $g_{\rm qb}$ between the qubit and the HBAR mode, where we assume that $g_{\rm qc} \gg g_{\rm qb}$. Finally, the two drives are described by the amplitude $\epsilon_{\rm p,d}$ with frequencies $\omega_{\rm p,d}$. The system is operated in the cavity-qubit dispersive regime $g_{\rm qc} \ll \vert \omega_{\rm r} - \omega_{\rm q} \vert$. We then consider the standard Schrieffer-Wolff transformation up to the first order in $g_{\rm qc}/ \vert \omega_{\rm r} - \omega_{\rm q} \vert$. Defining $\chi = g_{\rm qc}^2/(\omega_{\rm q} - \omega_{\rm r})$, the transformed Hamiltonian reads
\begin{equation}
\begin{aligned}
\frac{\hat{\mathcal{H}}^\prime}{\hbar} &= \frac{\tilde{\Omega}_{\rm q}}{2} \hat{\sigma}_{\rm z} + \omega_{\rm r} \hat{a}^\dagger \hat{a} + \omega_{\rm b} \hat{b}^\dagger \hat{b}  + \chi \hat{a}^\dagger \hat{a} \hat{\sigma}_{\rm z} + g_{\rm qb} (\hat{b}^\dagger \hat{\sigma}_- + \hat{b} \hat{\sigma}_+) 
- \frac{g_{\rm qb}}{g_{\rm qc}} \chi (\hat{a}^\dagger \hat{b}  + \hat{a} \hat{b}^\dagger ) \hat{\sigma}_{\rm z} \\
& + \sum_{j = p,d} \epsilon_{\rm j} (\hat{a} e^{i \omega_j t} + \hat{a}^\dagger e^{-i \omega_j t}) + \frac{g_{\rm qc}}{\omega_{\rm r}-\omega_{\rm q}} \sum_{j = p,d} \epsilon_{\rm j} (\hat{\sigma}_+ e^{i \omega_j t} + \hat{\sigma}_- e^{-i \omega_j t}).
\end{aligned}
\end{equation}
The Lamb-shifted qubit frequency is $\tilde{\Omega}_{\rm q} = \omega_{\rm q} + \chi$. Given the system's parameters, we will discard the qubit-mediated beam-splitter term $\frac{g_{\rm qb}}{g_{\rm qc}} \chi (\hat{a}^\dagger \hat{b} + \hat{a} \hat{b}^\dagger ) \hat{\sigma}_{\rm z}$, as it is several orders of magnitude smaller than $g_{\rm qb}$. Furthermore, in the two-tone spectroscopic setup, we can retain only the probe term for the cavity and only the drive term for the qubit. With such approximations, we have
\begin{equation}
\label{eq:Ham0}
\begin{aligned}
\frac{\hat{\mathcal{H}}^\prime}{\hbar} &= \frac{\tilde{\Omega}_{\rm q}}{2} \hat{\sigma}_{\rm z} + \omega_{\rm r} \hat{a}^\dagger \hat{a} + \omega_{\rm b} \hat{b}^\dagger \hat{b}  + \chi \hat{a}^\dagger \hat{a} \hat{\sigma}_{\rm z} + g_{\rm qb} (\hat{b}^\dagger \hat{\sigma}_- + \hat{b} \hat{\sigma}_+) \\
& +  \epsilon_{\rm p} (\hat{a} e^{i \omega_{\rm p} t} + \hat{a}^\dagger e^{-i \omega_{\rm p} t}) + \varepsilon_{\rm d} (\hat{\sigma}_+ e^{i \omega_{\rm d} t} + \hat{\sigma}_- e^{-i \omega_{\rm d} t}),
\end{aligned}
\end{equation}
where we have defined $\varepsilon_{\rm d}=\frac{g_{\rm qc} \epsilon_{\rm d}}{\omega_{\rm q}-\omega_{\rm r}}$. It is convenient to move to a frame co-rotating with the pump and the probe frequencies, for which the Hamiltonian in Eq. \eqref{eq:Ham0} reads
\begin{equation}
\label{eq:HamRot}
\begin{aligned}
\frac{\hat{\mathcal{H}}_{\rm rot}^{\prime}}{\hbar} &= -\frac{\Delta_{\rm q}}{2} \hat{\sigma}_{\rm z} +( -\Delta_{\rm r} + \chi) \hat{a}^\dagger \hat{a} - \Delta_{\rm b} \hat{b}^\dagger \hat{b}  + \chi \hat{a}^\dagger \hat{a} \hat{\sigma}_{\rm z} \\ &+ g_{\rm qb} (\hat{b}^\dagger \hat{\sigma}_- + \hat{b} \hat{\sigma}_+) 
 +  \epsilon_{\rm p} (\hat{a} + \hat{a}^\dagger) + \varepsilon_{\rm d} (\hat{\sigma}_+ + \hat{\sigma}_-),
\end{aligned}
\end{equation}
where $\Delta_{\rm b} = \omega_{\rm d} - \omega_{\rm b}$,  $\Delta_{\rm q} = \omega_{\rm d} - \tilde{\Omega}_{\rm q}$, and $\Delta_{\rm r} = \omega_{\rm p} - \omega_{\rm r} + \chi$. We have defined the readout cavity detuning $\Delta_{\rm r}$ to include the Lamb shift; however, this decision is arbitrary. The density matrix of the system $\rho$ has dynamics described by the master equation
\begin{equation}
\label{eq:MEQD}
    \partial_t \rho = - \frac{i}{\hbar} [\hat{\mathcal{H}}_{\rm rot}^{\prime}, \rho] + \kappa \mathcal{L}[\hat{a}] \rho + \gamma_{\rm{b}} \mathcal{L}[\hat{b}] \rho + \Gamma_1 \mathcal{L}[\hat{\sigma}_-] \rho + \frac{\Gamma_\phi}{2} \mathcal{L}[\hat{\sigma}_{\rm{z}}] \rho.
\end{equation}
Here $\kappa$ is the cavity decay, $\gamma_{\rm{b}}$ is the phonon decay, $\Gamma_1$ is the qubit population decay, and $\Gamma_\phi$ is the qubit dephasing. We have assumed zero temperature and ignored small corrections to the dissipator due to the Schrieffer-Wolff transformation.

\subsection{Mean field theory for the steady-state phonon population}

We can eliminate the microwave cavity mode in \eqref{eq:MEQD} following the procedure outlined in \cite{Gambetta2008quantumtrajectory}. The procedure corresponds to a displacement of the cavity mode conditioned on the qubit state, followed by a partial trace of the cavity mode under the assumption of no occupation of the cavity fluctuations. More details will be given in a future paper. The procedure yields the following master
\begin{equation}
\label{eq:MEQDEff}
    \partial_t \varrho = - \frac{i}{\hbar} [\hat{\mathcal{H}}_{\rm eff}, \varrho] + \gamma_{\rm{b}} \mathcal{L}[\hat{b}] \varrho + \Gamma_1 \mathcal{L}[\hat{\sigma}_-] \varrho + \frac{\tilde{\Gamma}_\phi}{2} \mathcal{L}[\hat{\sigma}_{\rm{z}}] \varrho,
\end{equation}
where $\varrho$  is the phonon-qubit density matrix, and $\tilde{\Gamma}_\phi$ is the qubit dephasing including the read-out cavity induced dephasing \cite{Gambetta2008quantumtrajectory}. The effective Hamiltonian reads
\begin{equation}
\frac{\hat{\mathcal{H}}_{{\rm{eff}}}}{\hbar} = - \Delta_{\rm{b}} \hat{b}^\dagger \hat{b} - \frac{\tilde{\Delta}_{\rm{q}}}{2} \hat{\sigma}_{\rm{z}} + g_{\rm{qb}}\left( \hat{b} \hat{\sigma}_+ +  \hat{b}^\dagger \hat{\sigma}_- \right) + \varepsilon_{\rm{d}} \left(\hat{\sigma}_+ + \hat{\sigma}_- \right).
\end{equation}
The qubit detuning $\tilde{\Delta}_q$  also considers the modification due to the read-out cavity.

The total qubit dephasing and the qubit detuning are given by
\begin{equation}
\begin{aligned}
\tilde{\Gamma}_\phi &= \Gamma_\phi + \Gamma_{\phi, \rm{cav}} (t), \\
\tilde{\Delta}_{\rm{q}} &= \Delta_{\rm{q}} - \omega_{\rm{q, cav}}(t),\\
\omega_{\rm{q, cav}}(t)(t) &= 2 \chi {\rm{Re}}[\alpha_{\rm{g}}(t) \alpha_e^*(t)], \\
 \Gamma_{\phi, \rm{cav}} (t) &=2 \chi {\rm{Im}}[\alpha_{\rm{g}}(t) \alpha_{\rm{e}}^*(t)].
\end{aligned}
\end{equation}
Here, $\alpha_{_{\rm{e(g)}}}(t)$ are the cavity amplitude if the qubit is in the excited (ground) state, given by
\begin{equation}
\label{eq:alpe0}
\begin{aligned}
\partial_t \alpha_{{\rm{e}}} &= \left[i (\Delta_{{\rm{r}}} - \chi) - \frac{\kappa}{2} \right] \alpha_{{\rm{e}}} - i \epsilon_{{\rm{p}}}, \\
\partial_t \alpha_{{\rm{g}}} &= \left[i (\Delta_{{\rm{c}}} + \chi) - \frac{\kappa}{2} \right] \alpha_{{\rm{g}}} - i \epsilon_{{\rm{p}}}.
\end{aligned}
\end{equation}
Since the cavity decay $\kappa$ is significantly larger than the other decays in the system, we consider for now on that the cavity is always in its steady state, such that the amplitudes $\alpha_{\rm{e(g)}}$ are given by
\begin{equation}
\begin{aligned}
\alpha_{\rm{e}} &= \frac{i \epsilon_{\rm{p}}}{i (\Delta_{{\rm{r}}} - \chi) - \frac{\kappa}{2}}, \\
\alpha_{\rm{g}} &= \frac{i \epsilon_{\rm{p}}}{i (\Delta_{{\rm{r}}} + \chi) - \frac{\kappa}{2}}.
\end{aligned}
\end{equation}

From the master equation \eqref{eq:MEQDEff}, we can then obtain the following equations for $\langle \hat{b} \rangle = b,$ $\langle \hat{\sigma}_- \rangle = s_-$, $\langle \hat{\sigma}_{\rm{z}} \rangle = s_{\rm{z}}$:
\begin{equation}
\label{eqs:MF}
\begin{aligned}
\partial_t b &= \left( i \Delta_{\rm{b}} - \frac{\gamma_{\rm{b}}}{2} \right) b(t) - i g_{\rm{qb}} s_-(t), \\
\partial_t s_- &= \left( i \tilde{\Delta}_{\rm{q}} - \tilde{\gamma}_2 \right) s_- (t) + i g_{\rm{qb}} b(t) s_{\rm{z}} (t) + i \varepsilon_{\rm{q}} s_{\rm z} (t), \\
\partial_t s_{\rm{z}} &= 2 i s_-(t) \left( g_{\rm{qb}} b^*(t) + \varepsilon_{\rm{d}} \right) - 2 i s_-^*(t) \left( g_{\rm{qb}} b(t) + \varepsilon_{\rm{d}} \right) - \gamma_1( s_{\rm{z}}(t) +1),
\end{aligned}
\end{equation}
which assume the mean field approximations $\langle \hat{b} \hat{\sigma}_{\rm{z}} \rangle \approx \langle \hat{b} \rangle \langle \hat{\sigma}_{\rm{z}} \rangle = b s_{\rm{z}}$ and $\langle \hat{b}^\dagger \hat{\sigma}_{-} \rangle \approx \langle \hat{b}^\dagger \rangle \langle \hat{\sigma}_{-} \rangle = b^* s_{-}$. The steady-state phonon number is given by $\bar{n}_b = \vert \bar{b} \vert^2$, where $\bar{b}$ is the steady-state of the phonon amplitude $b(t)$. Such steady-state is given by
\begin{equation}
\label{eq:phononumbqubit0}
\bar{n}_{\rm{b}} = \vert \bar{b} \vert^2 = \frac{g_{\rm{qb}}^2 \varepsilon_{\rm{d}}^2 \bar{s}_{\rm{z}}^2}{ \left(\Delta_{\rm{b}} \tilde{\gamma}_2 + \tilde{\Delta}_{\rm{q}} \frac{\gamma_{\rm{b}}}{2} \right)^2 + \left( \frac{\gamma_{\rm{b}} \tilde{\gamma}_2}{2} - \Delta_{\rm{b}} \tilde{\Delta}_{\rm{q}} - g_{\rm{qb}}^2 \bar{s}_{\rm{z}} \right)^2} ,
\end{equation}
where the steady-state $\bar{s}_{\rm{z}}$ reads

\begin{equation}
\label{eq:steadystatesz}
\bar{s}_{\rm z} = - \frac{\gamma_1}{\gamma_1 + 4 \frac{\tilde{\gamma}_2}{\tilde{\Delta}_{\rm{q}}^2 + {\tilde{\gamma}_2}^2}\vert g_{\rm{qb}} \bar{b} + \varepsilon_{\rm{d}} \vert^2}.
\end{equation}
The equation for the phonon number \eqref{eq:phononumbqubit0} is a non-linear equation for $\bar{b}$  that has to be solved numerically. The results for parameters in correspondence with the experiment are shown in Supplementary Fig.~ \ref{Fig:Mean_Field}, in which we can see a good agreement with the full master equation simulation.

\begin{figure}[h!]
\centering
\includegraphics[width = 0.4\textwidth]{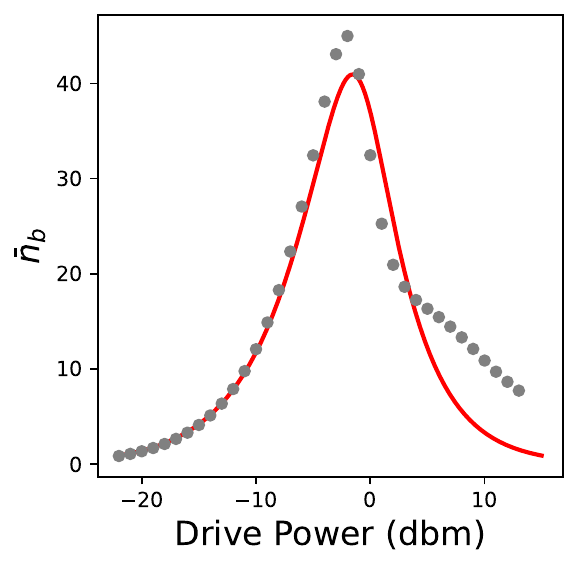}
\caption{\textbf{Steady-state mean-field phonon number as a function of the qubit drive power}. The red line depicts the result obtained with the mean-field solution of Eq.~\eqref{eq:phononumbqubit}, and the gray points depict the results from simulations of the full master equation \eqref{eq:MEQD}.  The parameters used are given in Table \ref{Table0} in correspondence with the experimental parameters of the main text.}
\label{Fig:Mean_Field}
\end{figure}

\begin{figure}[h!]
\centering
\includegraphics[width = 0.4\textwidth]{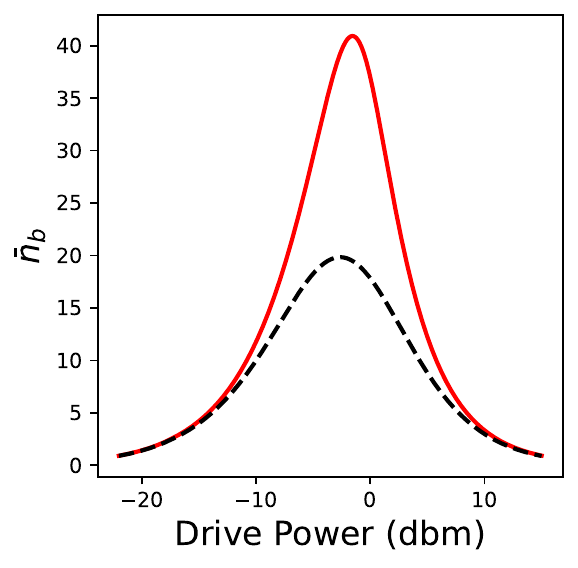}
\caption{\textbf{Steady-state mean-field phonon number as a function of the qubit drive power}. The red line depicts the result obtained with the mean-field solution of Eq.~\eqref{eq:phononumbqubit}, while the black dashed line depicts the result obtaining by discarding the phonon nonlinearity in eqs.~\eqref{eq:effectivepar} The parameters used are given in Table \ref{Table0} in correspondence with the experimental parameters of the main text.}
\label{Fig:Mean_Field2}
\end{figure}

We can further manipulate equation  \eqref{eq:phononumbqubit} as 

\begin{equation}
\label{eq:phononumbqubit}
\bar{n}_{\rm{b}} = \vert \bar{b} \vert^2 = \frac{g_{\rm{qb}}^2 \varepsilon_{\rm{d}}^2}{ \left(\Delta_{\rm{b}} \frac{\gamma_{\rm{eff}}}{2} + \Delta_{\rm{eff}} \frac{\gamma_{\rm{b}}}{2} \right)^2 + \left( \frac{\gamma_{\rm{b}} \gamma_{\rm{eff}}}{4} - \Delta_{\rm{b}} \Delta_{\rm{eff}} + g_{\rm{qb}}^2 \right)^2} ,
\end{equation}
where we have defined the effective decay $\gamma_{\rm{eff}}$ and detuning $\Delta_{\rm{eff}}$ as 
\begin{equation}
\label{eq:effectivepar}
\begin{aligned}
\frac{\gamma_{\rm{eff}}}{2}  &\equiv \frac{\tilde{\gamma}_2}{-\bar{s}_{\rm{z}}} = \frac{\tilde{\gamma}_2}{\gamma_1} \left(\gamma_1 + 4 \frac{\tilde{\gamma}_2}{\tilde{\Delta}_{\rm{q}}^2 + {\tilde{\gamma}_2}^2}\vert g_{\rm{qb}} \bar{b} + \varepsilon_{\rm{d}} \vert^2 \right), \\
\Delta_{\rm{eff}} &\equiv \frac{\tilde{\Delta}_{\rm{q}}}{-\bar{s}_{\rm{z}}} = \frac{\tilde{\Delta}_{\rm{q}}}{\gamma_1} \left(\gamma_1 + 4 \frac{\tilde{\gamma}_2}{\tilde{\Delta}_{\rm{q}}^2 + {\tilde{\gamma}_2}^2}\vert g_{\rm{qb}} \bar{b} + \varepsilon_{\rm{d}} \vert^2 \right).
\end{aligned}
\end{equation}
From such a mean-field perspective, the steady-state of the phonon mode behaves similarly as if it would be coupled to a coherently driven cavity with the nonlinear decays and detunings given by Eq.~ \eqref{eq:effectivepar}. We notice that as the power is increased, the effective decay $\gamma_{\rm{eff}}$ increases, a consequence of the qubit power broadening. Such an effect yields the eventual decoupling between the qubit and the phonon, similar to an inverse Purcell effect. Nevertheless, such a power broadening is not the only physical mechanism that plays a role in the build-up of the phonon steady-state. In fact, both the effective decay and detuning depend on the phonon amplitude $\bar{b}$, a non-linearity that can not be discarded for the experimental parameters. To show this effect, we plot in Fig. \ref{Fig:Mean_Field2} the mean-field solution for $n_{\rm{b}}$ (red curve) and the corresponding curve by setting $g_{\rm{qb}}=0$ in Eqs.~\eqref{eq:effectivepar} (dashed black curve). The latter still captures the inverse Purcell effect due to the qubit power broadening but yields half of the maximum phonon occupancy, indicating the importance of the phonon nonlinearity stemming from the phonon-qubit coupling and the intrinsic two-level nature of the qubit.

\subsection{Phonon mode linewidth}
\label{sec:phononlinewidth}

The semi-classical equations describing the system dynamics \eqref{eqs:MF} can also be used to obtain information about the phonon mode response in the steady-state, in particular, its effective linewidth. To compute the effective phonon linewidth in the steady-state, we numerically solve Eq.~\eqref{eqs:MF} in time-domain, which gives the time-dependent phonon amplitude $b(t)$. The phonon spectrum is then given by $\vert b[\omega] \vert^2$, where $b[\omega]$ is the Fourier transform of $b(t)$. Such a spectrum is a Lorentzian in the frequency domain, with a linewidth that we refer as the Phonon linewidth.

In Supplementary Fig.~\ref{Plot_Linewidth} we shown in (A) the steady-state phonon population as a function of the drive amplitude for the parameters of the experiment and in (B) the corresponding phonon linewdith obtained by fitting the phonon spectrum obtained numerically to a Lorentzian. In (C) and (D) we show two examples of phonon spectrum. We notice that as the drive approaches the upper threshold, the phonon linewidth goes below its intrinsic value, pointing to an amplification of the phonon mode. Such an effect would not be possible only with a decoupling between qubit and phonon, and it is intrinsically related to the non-linear character of the Jaynes-Cummings interaction. Furthermore, the reduction of the effect linewidth is what enables the build-up of such a large steady-state phonon population.

\begin{figure*}[b!]
    \centering
    \includegraphics[width = 0.75\textwidth]{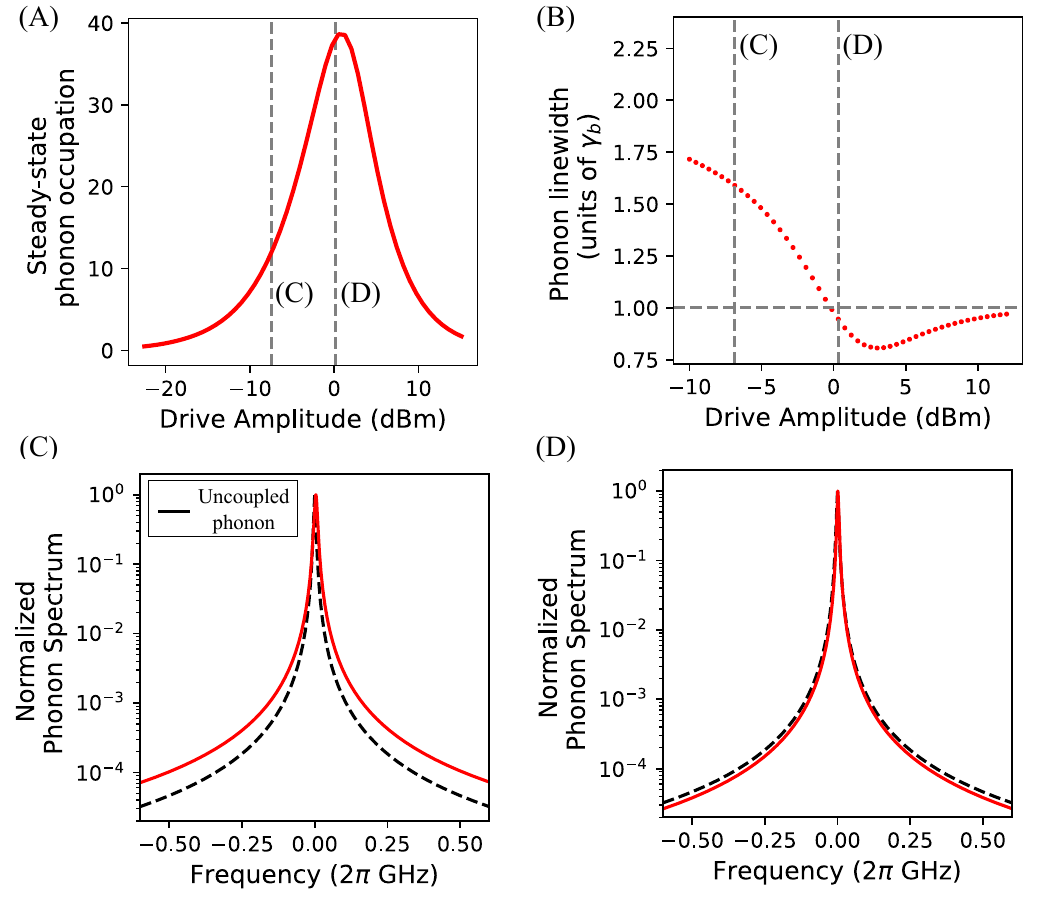}
    \caption{\textbf{Linewidth narrowing of the HBAR mode} (A) Steady-state phonon occupation for the system in the manuscript. (B) Effective phonon linewidth extracted from a Lorentzian fit of the phonon spectrum. (C,D) phonon spectra at two representative drive amplitudes showing the linewidth narrowing with power, the black dashed line depicts the spectrum for a phonon mode uncoupled from the two-level system. Parameters in correspondence with the experiment.}
    \label{Plot_Linewidth}
\end{figure*}

\subsection{Qubit Two-Tone Spectroscopy}
\label{Sec:ng}

Following the arguments presented in \cite{gambetta2006qubit}, ignoring the qubit-phonon coupling, in the dispersive regime, ${\rm{arg}}[\langle \hat{a} (t) \rangle]$  is directly related to the qubit population $\langle \hat{\sigma}_z(t) \rangle$. By recording the phase of the readout resonator, one can then obtain the qubit absorption spectrum
\begin{equation}
\label{eq:qubitabsspect}
S(\omega) = \frac{1}{2 \pi} \int_{- \infty} ^\infty  dt e^ {i \omega t} \langle \hat{\sigma}_-(t) \hat{\sigma}_+ (0) \rangle_s,
\end{equation}
where $\langle \cdot \rangle_s$ indicates that the expectation value is taken in the steady state.

It was shown that for a qubit-cavity system, the qubit absorption spectrum is given by \cite{gambetta2006qubit}
\begin{equation}
\label{eq:barqubitspectrum}
S(\omega) = \frac{1}{\pi} \sum_{j = 0}^\infty \frac{1}{j!} {\rm{Re}} \left( \frac{(-A)^j e^A}{\Gamma^{(j)}_{\rm q}/2 - i (\omega - \Omega^{(j)}_{\rm q})} \right) \equiv \sum_{j = 0}^{\infty} S_{j}(\omega),
\end{equation}
with
\begin{equation}
\begin{aligned}
    A &= D_{ss} \bigg( \frac{\kappa/2 -2i\chi}{\kappa/2 + 2i\chi} \bigg), \\
    B &= \chi (\bar{n}_e + \bar{n}_g - D_{ss}), \\
    D_{ss} &= \frac{2 \chi^2 (\bar{n}_e + \bar{n}_g)}{(\kappa/2)^2 + 2\chi^2}, \\
    \bar{n}_e &= \frac{\bar{n}_g (\kappa/2)^2}{(\kappa/2)^2 + (2\chi)^2}, \\
    \Gamma^{(j)}_{\rm q} &= 2\gamma_{\rm q} + \kappa(j + D_{ss}), \\
    \omega^{(j)}_{\rm q} &= \tilde{\omega}_q + B + 2j\chi.
\end{aligned}
\end{equation}
In the above equations, $\omega^{(j)}_{\rm q}$ and $\Gamma^{(j)}_{\rm q}$ are the frequency and linewidth of the qubit with the readout resonator in the state $\vert j \rangle$, respectively. The intrinsic qubit linewidth, with the readout resonator in its ground state, is given by $\gamma_{\rm q} = \Gamma_1/2 + \Gamma_{\phi}$, where $\Gamma_1$ is the longitudinal relaxation rate, and $\Gamma_{\phi}$ is the pure dephasing rate. The qubit frequency $\omega^{(0)}_{\rm q}$ is ac Stark shifted by $B$ from its intrinsic value $\omega_{\rm q}$. We have also assumed that the readout drive is on resonance with the readout resonator, i.e. $\Delta_{\rm r} = 0$, and the readout resonator has a full-width half-maximum linewidth $\kappa$. In the limit $\chi \sim \kappa$ and with $\Delta_{\rm r} = 0$, the components $S_j(\omega)$ have non-Lorentzian lineshapes and can even be negative. The sum of these individual components can result in an asymmetry of the qubit spectrum, as seen in Supplementary Fig.~\ref{SI_Fig_ng}(a).

We perform spectroscopy of the qubit by monitoring the transmission coefficient $S_{21}$ of the readout resonator as a function of the qubit drive frequency $\omega_{\rm d}$. The probe tone was fixed at the Stark-shifted readout resonator frequency with the qubit in its ground state $\omega^g_{\rm c}/2\pi = 4.91$ GHz, such that $\Delta_{\rm r} = 0$ and held at a constant power $\mathcal{P}_{d} = -25$ dBm, set at room temperature. The probe line has a total of 108 dB of attenuation, ensuring the average number of photons in the probe mode on average is much less than one.

To fit the measured spectrum, we use the expression,

\begin{equation} 
    \vert S_{21} \vert = \mathcal{A} \sum_{j = 0}^{10} S_{j}(\omega) + \mathcal{C},
\label{S21_Fit}
\end{equation}
where $\mathcal{A}$ is a conversation factor between $S_{j}(\omega)$ and $\vert S_{21} \vert$ and $\mathcal{C}$ is a constant offset of the spectrum. The value of the Fock basis was truncated to $j = 10$, and the linewidth of the readout resonator was independently determined and fixed; see Table.~\ref{Table0}. The fitting parameters include the intrinsic qubit frequency $\omega_{\rm q}$, the power broadened qubit linewidth $\Gamma_{q}(\mathcal{P}_{\rm d})$, where $\mathcal{P}_{\rm d}$ is the qubit drive power, the qubit dispersive shift $\chi$, the probe mode occupancy with the qubit in its ground state $\bar{n}_{\rm g}$, and conversion factor $\mathcal{A}$, and the constant offset $\mathcal{C}$.

\begin{figure*}[t]
    \centering
    \includegraphics[width = 0.9\textwidth]{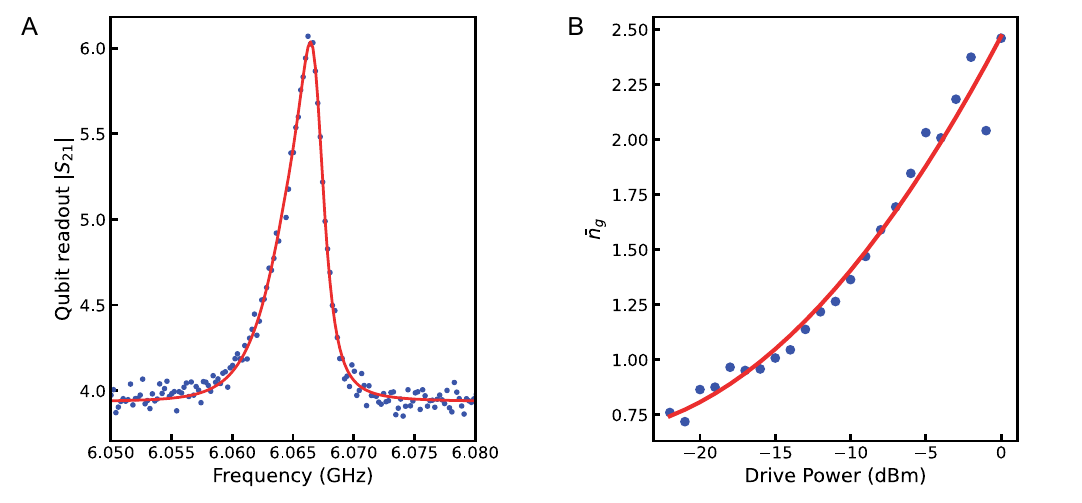}
    \caption{\textbf{Two-tone qubit spectroscopy measurements.} (A) Experimental two-tone qubit spectrum (blue) and the qubit spectrum determined from a fit to Eq.~\ref{S21_Fit} (red). The qubit drive power at room temperature was set to -20.0 dBm. (B) Extracted value of the readout resonator photon population as a function of qubit drive power.}
    \label{SI_Fig_ng}
\end{figure*}

An example spectrum and its fit are shown in Supplementary Fig.~\ref{SI_Fig_ng}(a). First, the value of the dispersive shift was determined to be $\chi/2\pi = -1.2 \pm 0.2$ MHz, which agreed with our designed value. This value was then fixed, and the data was fit for all qubit drive powers to determine the remaining values. We found that the ground state readout photon population depended on the qubit drive power. This is likely due to the heating of the silicon substrate because the qubit drive tone was applied via the readout resonator. The value of $\bar{n}_{\rm g}(\mathcal{P}_{\rm d})$ is shown in Supplementary Fig.~\ref{SI_Fig_ng}(b). The zero power qubit linewidth was also extracted by extrapolating the measured power-broadened qubit linewidth to zero power and was determined to be $\gamma_{\rm q}(0)/2\pi = 0.420 \pm 0.04$ MHz, which agrees with our $\gamma_{\rm q}/2\pi = \Gamma_1 / 2 + \Gamma_{\phi} \geq 1/2T_1$ limit determined using a time-domain measurement with $T_1 = 180 \pm 30$ ns. The comparison between our spectroscopy and time domain measurements indicates that our qubit decay is dominated by decoherence, and therefore we have ignored intrinsic dephasing in our model since qubit dephasing will be dominated by power-induced dephasing induced by the qubit drive tone.

\subsection{Master Equation Simulations}

\subsection*{Qubit Spectroscopy}

\begin{figure*}[t]
    \centering
    \includegraphics[width = 0.9\textwidth]{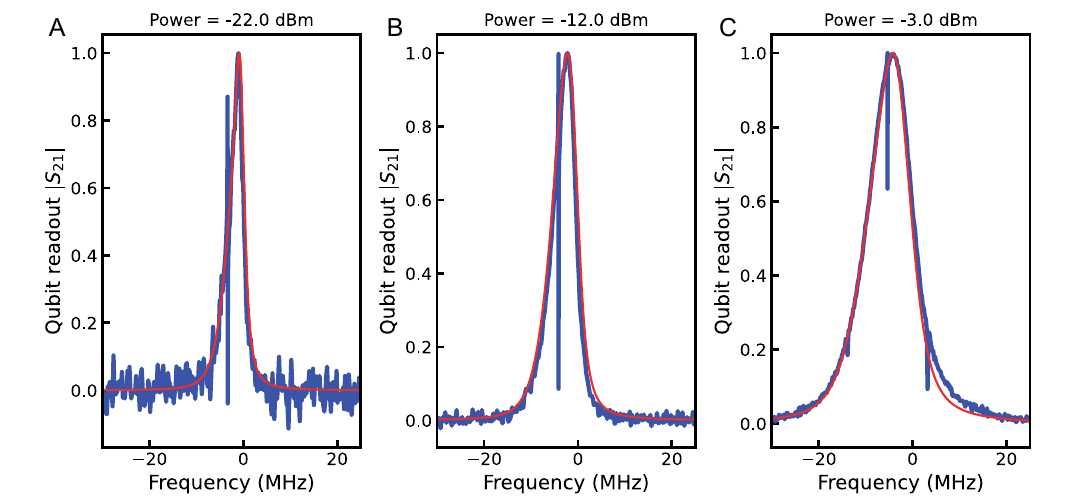}
    \caption{\textbf{Two-tone qubit spectroscopy measurements.} (A) Experimental two-tone qubit spectrum (blue) overlapped with the qubit spectrum determined from the master equation simulation (red). The frequency is defined as the detuning from the bare qubit frequency. The qubit drive power at room temperature was set to -12.0 dBm. (B) Qubit spectrum and master equation simulation for a qubit drive power of -6.0 dBm. (C) Qubit spectrum and master equation simulation for a qubit drive power of -3.0 dBm. The feature at $\sim 0$ MHz is an additional HBAR mode, one free-spectral range from the HBAR mode of interest for this work. The only parameter that was varied within the simulation was the room-temperature value of the qubit drive.}
    \label{SI_Fig_Qubit_Cal}
\end{figure*}

We simulate the dynamics of the master equation Eq.~\ref{eq:MEQD} using the Python package Qutip \cite{johansson2012qutip}. We first compare the measured qubit spectrum without the phonon mode to the Qutip steady-state simulations. To begin, we must consider the non-zero photon population of the readout resonator. The finite population results in an asymmetry in the qubit spectrum, as well as additional measurement-induced dephasing. To include the finite readout population in our simulation, the readout drive $\epsilon_{\rm p}$ was set such that the average population $\langle \hat{a}^{\dagger} \hat{a} \rangle = \bar{n}_{\rm g}(\mathcal{P}_{\rm d})$. This ensured that for each qubit drive power, the readout resonator had the appropriate number of steady-state photons.

To account for qubit power-broadening, the qubit drive power in the simulation had to be calibrated. As stated above, $\varepsilon_{\rm d} =  \frac{g_{\rm qc} \epsilon_{\rm d}}{\omega_{\rm r}-\Omega_{\rm q}}$, where $\epsilon_{\rm d} = \sqrt{\kappa_{\rm ext}}\sqrt{\mathcal{P}_{\rm d} / \hbar \omega_{\rm d}}$, $\kappa_{\rm ext}$ is the external coupling rate to the readout resonator, and $\mathcal{P}_{\rm d}$ is the drive power in Watts at the coupling port of the readout resonator. Therefore, the drive coefficient $\varepsilon_{\rm d}$ can be written in the form
\begin{equation}
    \varepsilon_{\rm d} = \sqrt{10^{(\mathcal{P}_{\rm RT} + \delta)/10}}
\end{equation}
where $\delta$ calibrates the room-temperature power to the corresponding value of $\mathcal{P}_{\rm d}$ accounting for all losses and multiplicative factors. The value of $\delta$ was determined by matching the simulation to the power-broadened qubit spectrum at multiple qubit drive powers. A set of qubit spectra is shown in Supplementary Fig.~\ref{SI_Fig_Qubit_Cal} where it can be seen that the master equation simulation is in excellent agreement capturing the qubit asymmetry at low power, and the power-broadened qubit spectrum at higher drive powers. It should be noted the measured transmission signal $S_{21}$ is proportional to the qubit population $\langle \hat{\sigma}_{\rm z} \rangle$ only if the appropriate signal quadrature is measured. Here, this was determined by calculating the rotation angle that minimized the signal in the out-of-phase quadrature. To confirm, we compared our single-trace two-tone data with a direct measurement of the readout resonator frequency at several qubit drive powers. At the highest drive powers, we observed a slight deviation resulting from signal mixing into the out-of-phase quadrature. However, this was at the drive powers above the self-quenching threshold and resulted in a slight mismatch between our experimental and simulated two-tone traces. 

\subsection*{Gated Ringdown}

\begin{figure*}[t]
    \centering
    \includegraphics[width = 0.9\textwidth]{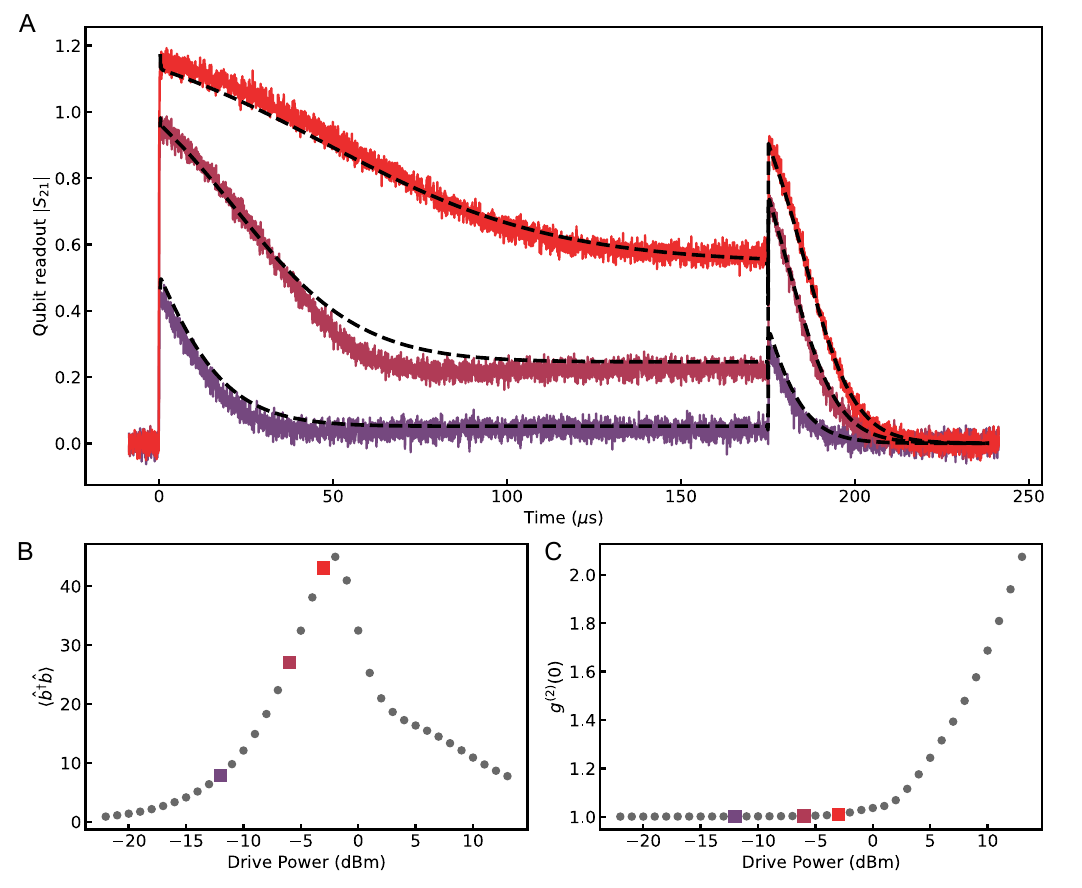}
    \caption{\textbf{Gated two-tone qubit spectroscopy.} (A) Experimental gated two-tone measurements for qubit drive powers -12.0, -6.0, -3.0 dBm. The master equation simulations at the same drive power are plotted as dashed black lines. The simulation has an excellent agreement for the ringdown data for all powers and a slight deviation from the ring-up data. This could be attributed to higher-order non-linearities we are not considering within our simple master equation model. However, the agreement between the experiment and simulation is good for all powers. (B) Simulated phonon state population $\langle \hat{b}^{\dagger}\hat{b} \rangle$ as a function of qubit drive power. (C) Simulated second-order phonon correlations $g^{(2)}(0)$ as a function of qubit drive power.}
    \label{SI_Fig_Gated}
\end{figure*}

Following the calibration of the qubit spectrum, the phonon mode was included in the master equation simulation. The value of the coupling rate $g_{\rm qb}$ and the phonon linewidth $\gamma_{\rm b}$ were determined by performing a fit to the time domain gated two-tone measurements. The fit was determined by fitting both the ringdown (as shown in the main text) and also the ring-up of the qubit. Here, the device was allowed to thermalize before the gated-ringdown measurement, weakly probing the readout resonator continually. At the time $t=0$, the qubit drive is switched on for $175 \mu$s, allowing the phonon moded to ring-up. At $t=175 \mu$s, the qubit drive is switched off, and the phonon ringdown is observed, as described in the main text. The best-fit values were compared at multiple drive powers; see Supplementary Fig.~\ref{SI_Fig_Gated}(a). We extracted a value of the qubit phonon coupling of $g_{\rm qb}/2\pi = 162$ kHz and a phonon linewidth of $\gamma_{\rm b}/2 \pi = 6.81$ kHz. From this set of simulations, the steady-state phonon population and phonon statistics can be estimated for multiple drive powers. We observe a good agreement between the numerical simulation and the ring-up data and an excellent agreement between the ringdown data for all powers. Deviations in the ring-up simulations likely result from higher-order nonlinearities we are not considering in our model.

The master equation simulations directly calculate the time dynamics of $\langle \hat{\sigma}_+ \hat{\sigma}_- \rangle$, this value has been scaled by the same constant factor for all powers to give a direct comparison to our two-tone measurement. Since we expect the variation in the transmission coefficient measured in the two-tone measurement to be proportional to $\langle \hat{\sigma}_+ \hat{\sigma}_- \rangle$, see Eq.~\ref{eq:Ham0} and  Ref.~\cite{gambetta2006qubit}. From our simulation, we calculate the phonon population and second-order correlation function for multiple drive powers, shown in Supplementary Fig.~\ref{SI_Fig_Gated}(b,c). For low drive powers, the phonon statistics are described by a coherent state, $g^{(2)}(0) \approx 1.0$. However, as discussed in the main text, the phonon amplitude exhibits self-quenching above a given upper-threshold power. This can be seen in both the phonon population, as a rapid decrease in the populations, and in the phonon statistics as $g^{(2)}(0) > 1.0$ for drive powers above the self-quenching threshold.

\begin{table*}[t!]
\begin{center}
\caption{Symbols and parameters
\label{Table0}}
\begin{tabular}{ |c|c| } 
\hline
\textbf{Parameter} & \textbf{Symbol} \\
\hline
Microwave mode frequency & $\omega_{\rm c} = 2 \pi \times 4.910$ GHz \\
\hline
Microwave mode decay & $\kappa = 2 \pi \times 2.897$ MHz \\
\hline
Phonon mode frequency & $\omega_{\rm b} = 2 \pi \times 6.064$ GHz \\
\hline
Phonon mode decay & $\gamma_{\rm b} = 2 \pi \times 6.81$ kHz \\
\hline
Qubit frequency & $\Omega_{\rm q} = 2 \pi \times 6.067$ GHz \\
\hline
Qubit energy relaxation rate & $\Gamma_1 = 2 \pi \times 0.840$ MHz \\
\hline
Qubit phase relaxation rate & $\Gamma_\phi < 2 \pi \times 0.08$ MHz \\
\hline
Qubit anharmonicity & $\alpha = -2 \pi \times 260.0$ MHz  \\
\hline
Qubit-phonon coupling & $g_{\rm qb} = 2 \pi \times 162$ kHz  \\
\hline
Dispersive cavity-qubit coupling & $\chi = -2 \pi \times 1.2$ MHz  \\
\hline
Lamb-shifted qubit frequency & $\tilde{\Omega}_{\rm q} = \Omega_{\rm q} - \chi$  \\
\hline
Qubit drive frequency & $\omega_{\rm d}$, varied around $\Omega_q$ \\ 
\hline
Cavity probe frequency & $\omega_{\rm p}$, set at the Stark-shifted cavity frequency \\
\hline
\end{tabular}
\end{center}
\end{table*}

\section{Supplementary Note 3: Simplified Model}

\subsection{Considerations of Qubit Anharmonicity}

We modeled our system using a two-level approximation; however, under strong drives, it is important to consider the higher levels of the transmon qubit. We can instead model the qubit as a Kerr oscillator with a large anharmonicity $\alpha$. Therefore, we can re-write the Hamiltonian in the form
\begin{equation}
    \frac{\hat{\mathcal{H}}_{\rm mod}}{\hbar} = -\Delta_{\rm q} \hat{c}^{\dagger}\hat{c} + \frac{\alpha}{2}\hat{c}^{\dagger}\hat{c}^{\dagger}\hat{c}\hat{c} - \Delta_{\rm b} \hat{b}^\dagger \hat{b} + g_{\rm qb} (\hat{b}^\dagger \hat{c} + \hat{b} \hat{c}^{\dagger}) + \varepsilon_{\rm d} (\hat{c}^{\dagger} + \hat{c}).
\end{equation}
Where we have replaced the two-level raising and lowering operators with bosonic raising and lowering operators $\hat{c}^{(\dagger)}$. If we calculate the number of steady-state phonons comparing the Kerr oscillator model to the two-level system model, we find that at powers below and near the upper threshold, the population is slightly larger in the Kerr oscillator model; see Supplementary Fig.~\ref{SI_Fig_Anharm}. There exists a discrepancy at high drive powers; however, this was not observed within the experiment since we likely could not apply a large enough drive without heating the sample. Thus, we are confident that modeling our experiment as a two-level system is accurate, especially for calculating the upper threshold. Moreover, this was required for numerical efficiency. The two-level system simulation in Supplementary Fig.~\ref{SI_Fig_Anharm} ran for approximately 20 minutes, whereas considering only three levels of the Kerr oscillator increased the simulation time to approximately 10 hours.

\begin{figure*}[t!]
    \centering
    \includegraphics[width = 0.55\textwidth]{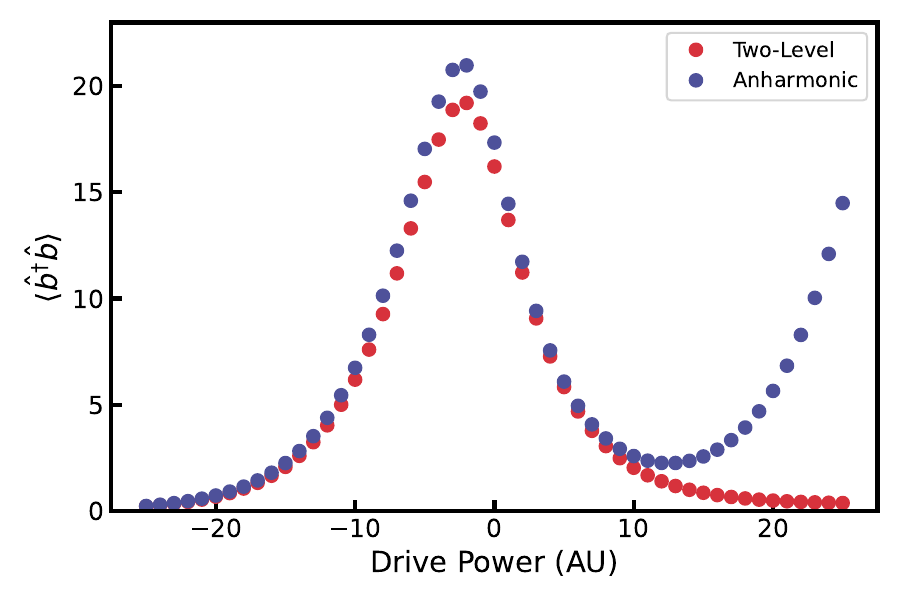}
    \caption{\textbf{Two-Level versus Anharmonic Oscillator Simulations.} The simulated phonon population when considering the transmon qubit as a two-level system versus considering three levels of the Kerr oscillator.}
    \label{SI_Fig_Anharm}
\end{figure*}

For numerical efficiency, we have constructed a simplified model where we have removed the coupling to the readout resonator. Studying Eq.~\ref{eq:Ham0}, the readout resonator broadens the qubit and causes a constant frequency offset due to steady-state photons in the cavity. The Hamiltonian of our simplified model is given by
\begin{equation}
    \frac{\hat{\mathcal{H}}_{\rm mod}}{\hbar} = -\frac{\Delta_{\rm q}}{2} \hat{\sigma}_{\rm z} - \Delta_{\rm b} \hat{b}^\dagger \hat{b} + g_{\rm qb} (\hat{b}^\dagger \hat{\sigma}_- + \hat{b} \hat{\sigma}_+) + \varepsilon_{\rm d} (\hat{\sigma}_+ + \hat{\sigma}_-),
\end{equation}
and we choose to have the qubit and phonon resonant for the simulation. Moreover, we have increased the intrinsic qubit linewidth $\Gamma_{\rm q}/2\pi = 1.5$ MHz to account for the lack of readout resonator broadening. We also increased the phonon linewidth $\gamma_{\rm b}/2\pi =  25$ kHz to reduce the size of the Hilbert space for numerical efficiency.

\subsection{Phonon Phase Coherence}

First, we aim to investigate if the drive tone seeds the phase of the phonon mode. In the model described in Ref.~\cite{ashhab2009single} the phase of the phonon mode is random and results in a distribution in the IQ plane. However, if we examine the phase space distribution of the phonon mode in our model, we find that the phase of the drive tone instead sets the phase; see Supplementary Fig.~\ref{SI_Fig_Phase}. This results from the weak hybridization between the qubit and phonon mode. When the drive is switched on, the weak hybridization enables the seeding of the phase of the phonon mode. This is followed by the stimulated emission of phonons, which produces a large coherent state with a set phase. The ability to set the phase of the phonon mode is important for applications such as the generation of Schr\"{o}dinger cat states.

\begin{figure*}[t]
    \centering
    \includegraphics[width = 0.95\textwidth]{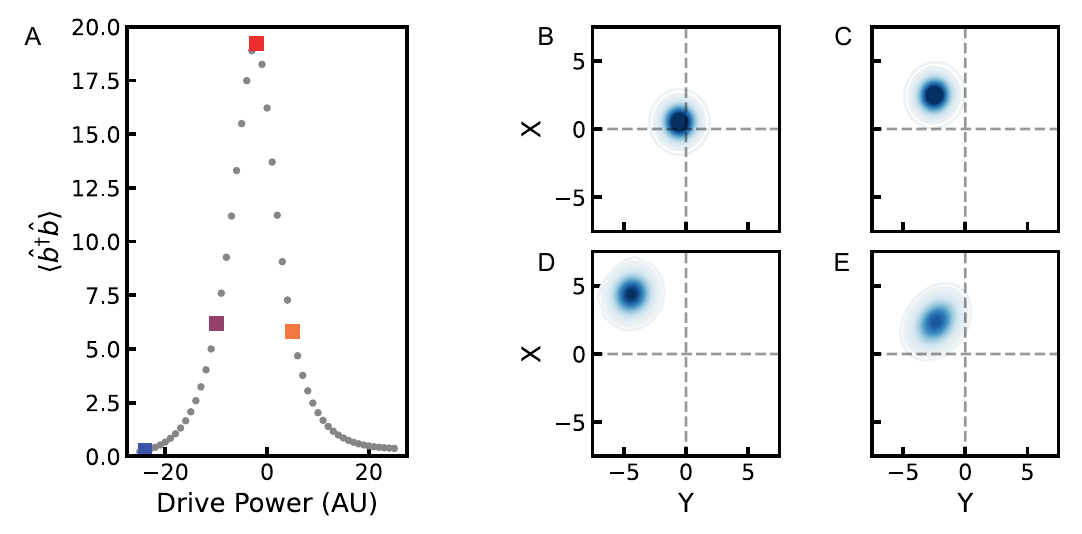}
    \caption{\textbf{Phonon Phase Seeding.} (A) Phonon population as a function of qubit drive power. Colored points correspond to the Wigner plots (B-E). (B-E) Wigner distribution plots corresponding to the points in (A). The phase of the coherent state is set by the drive phase; here, the drive phase was $\pi/4$. The plots are shown starting from the lowest power (B) to the largest power (E).}
    \label{SI_Fig_Phase}
\end{figure*}

\subsection{Phonon Mode Anharmonicity}

Finally, the coupling between the qubit and phonon mode raises the question of how much anharmonicity is inherited by the phonon mode. In the strong coupling limit, the phonon mode will inherit half of the qubit anharmonicity and, thus, would be unable to populate the phonon mode since its Hilbert space is truncated to a single excitation. From the transparency window measurement, we observe little frequency shift of the phonon mode, suggesting the mode is well approximated as a harmonic oscillator and \textit{not} a Duffing oscillator. However, again, we can utilize our simplified model to explore the inherited anharmonicity by directly driving the phonon mode in the simulation. We find that at a drive strength that generates, on average, fifty phonons, similar to what was observed in the experiment, the phonon mode has a negligible frequency shift. Suggesting that in the weak hybridization limit, the phonon mode can be treated to a high degree as a linear harmonic oscillator.

\clearpage

\end{document}